\documentclass[aps,prc,twocolumn,showpacs]{revtex4}
\usepackage{graphicx}
\usepackage{hyperref}
\usepackage{latexsym}
\usepackage{bm}

\newcommand{\bra}{\langle}
\newcommand{\ket}{\rangle}

\begin{document}
    
\bibliographystyle{apsrev}    

\title{The $\Lambda$-$\Lambda$ Interaction and $^{\ \; 6}_{\Lambda\Lambda}$He}

\author{I.\ R.\ Afnan}
\email[E-mail:]{Iraj.Afnan@Flinders.edu.au}
\affiliation{School of Chemistry, Physics and Earth Sciences\\
Flinders University, GPO Box 2100, Adelaide 5001, Australia}
\author {B.\ F.\ Gibson}
\email[E-mail:]{gibson@paths.lanl.gov}
\affiliation{Theoretical Division, Los Alamos National Laboratory\\ 
Los Alamos, New Mexico 87545, U.S.A.}

\begin{abstract}
An OBE potential model for the $^1$S$_0$ $S=-2$ interaction is analyzed 
with emphasis on the role of coupling between the $\Lambda\Lambda$,
N$\Xi$, and $\Sigma\Sigma$ channels.  Singlet scalar exchange, an 
approximation to two-pion exchange, is significant in all channels;
surprisingly, the one-pion exchange component is almost negligible.   
The size of the channel coupling as a function of the overall strength 
of the OBE model potential is examined.  Implications of the 
analysis for the binding energy of $^{\ \ 6}_{\Lambda\Lambda}$He 
are considered; the new experimental datum may suggest a consistency 
between the extracted $\Lambda\Lambda$ matrix element and the relation 
implied by $SU(3)$ among OBE baryon-baryon interactions. \\
\end{abstract} 

\maketitle


A recent $^{\ \ 6}_{\Lambda\Lambda}$He binding energy measurement\cite{T01}
yielding a $\Lambda\Lambda$ separation energy of
\begin{eqnarray}
\Delta B_{\Lambda\Lambda} &=&
B_{\Lambda\Lambda}(^{\ \ 6}_{\Lambda\Lambda}\mbox{He}) -
2B_\Lambda(^{\,5}_{\Lambda}\mbox{He}) \nonumber \\
 &=& 1.01 \pm 0.20 ^{+0.18}_{-0.11}\ \mbox{MeV}
                                             \label{eq:1.1}
\end{eqnarray}
suggests that the effective $\Lambda\Lambda$ interaction is considerably 
weaker than that inferred from the earlier measurement ($\approx 4.7$ MeV)
reported by Prowse\cite{P66}.  We examine the implication of this new 
measurement within the framework of One Boson Exchange (OBE) models that 
employ $SU(3)$ symmetry to determine the baryon-baryon strangeness $S= -2$
interaction.

If one assumes flavour $SU(3)$ is a good symmetry, then one can express 
the matrix elements of an OBE potential in terms of the irreducible 
representations of $8 \otimes 8$ as
\begin{eqnarray}
\bra nn|V|nn\ket &=& V_{27} \nonumber \\
\bra \Lambda N|V|\Lambda N\ket &=& \frac{36}{40}\,V_{27} 
                            + \frac{4}{40}\,V_{8_s} \label{eq:1.2}\\
\bra\Lambda\Lambda|V|\Lambda\Lambda\ket &=&
        \frac{27}{40}\,V_{27} + \frac{8}{40}\,V_{8_s} 
        + \frac{5}{40}\,V_{1}\ . \nonumber
\end{eqnarray}
Considering that $V_{8_s}$ and $V_{1}$ are repulsive while $V_{27}$ is 
attractive\cite{RMS92}, we may conclude that
\begin{equation}
|\bra\ V_{nn}\ \ket | > |\bra\ V_{\Lambda N}\ \ket | 
> |\bra\ V_{\Lambda\Lambda}\ \ket | \ .              \label{eq:1.3}
\end{equation}
From the three earlier measurements of $\Lambda\Lambda$ hypernuclei 
binding energies ($^{\ \; 6}_{\Lambda\Lambda}$He\cite{P66}, 
$^{\, 10}_{\Lambda\Lambda}$Be\cite{D63,D89}, and 
$^{\, 13}_{\Lambda\Lambda}$B\cite{A90,A91,D91}) which implied that the 
$\Lambda\Lambda$ matrix element $|\bra\Lambda \Lambda|V|\Lambda\Lambda\ket| $ 
was $\approx$ 4-5 MeV, it was suggested that the breaking of $SU(3)$ 
symmetry and the coupling between the $\Lambda\Lambda$, N$\Xi$, and
$\Sigma\Sigma$ channels in the $^1$S$_0$ partial wave could bridge the 
gap between experiment 
($|\bra V_{nn}\ket | > |\bra V_{\Lambda\Lambda}\ket | > 
  |\bra V_{\Lambda N}\ket |$) 
and the $SU(3)$ expectations expressed in Eq.~(\ref{eq:1.3}).

To examine this issue and the implications of the new experimental result, 
we consider the Nijmegen OBE potential Model $D$~\cite{NRS77}.  If we 
require all coupling constants be determined by the $SU(3)$ rotation of 
those parameters as fixed in the nucleon-nucleon (NN) and hypron-nucleon 
(YN) sectors, then the only free parameters are those of the short range 
component of the interaction.  These we vary within the constraint that 
the long range part of the potential is predominantly OBE in origin.  
This allows us to examine the $\Lambda\Lambda$ matrix element as 
a function of the strength of the $\Lambda\Lambda$ interaction and the 
importance of the coupling of the $\Lambda\Lambda$ channel to the N$\Xi$ 
and $\Sigma\Sigma$ channels. 

To perform an $SU(3)$ rotation on an OBE potential defined in the 
$S=0,-1$ sectors, one writes the Lagrangian in terms of the baryon 
octet coupled with the mesons which are either a singlet or a member of 
an octet. If the interaction is taken to be of the Yukawa type, then the 
interaction Lagrangian takes the form\cite{S63}
\begin{eqnarray}
L_{int} &=& -\left\{ g_s[B^{\dag} B]_s M_s + g_{8_1}[B^{\dag} B]_{8_1} 
M_8\right. \nonumber \\
&&\left.\qquad+  g_{8_2}[B^{\dag} B]_{8_2} M_8 \right\}\  ,   \label{eq:2.1}
\end{eqnarray}
where $B$ and $M$ are the baryon and meson field operators.  In writing 
this Lagrangian, which is a scalar, the initial and final baryons are
coupled to either a flavour singlet or an octet.  Because there exist 
two irreducible octet representations, one needs a different coupling 
constant for each of the representations.  That is, one has one coupling 
constant for each singlet meson $g_{s}$, and two coupling constants
$g_{\{8_{1}\}}$ and $g_{\{8_{2}\}}$ for each meson octet.  These coupling 
constants can then be determined by fitting the NN and YN experimental 
data.

The Nijmegen Model $D$ potential\cite{NRS77} postulates for the exchanged 
mesons the pseudoscalar octet $\{\pi,\eta,\eta',K\}$, the vector octet 
$\{\rho,\phi,\omega,K^{*}\}$, and a scalar meson $\{\varepsilon\}$.  The 
masses of the mesons and baryons are taken from experiment, while the 
coupling constants are adjusted to fit the data in the $S=0,-1$ sectors;
a hard core models the short range interaction.  This, in 
principle, determines the long range part of the potential which should 
be described in terms of meson-baryon degrees of freedom.  These same 
coupling constants can be used to construct an OBE potential for 
$S\le -2$.  Flavour $SU(3)$ is explicitly broken as a result of using
physical masses for the baryons and mesons and the difference in 
the short range properties of the potential as we proceed from the 
$S=0$ to the $S=-1$ and $S=-2$ channels.  

Such a procedure was followed by Carr {\it et~al}.\cite{C97}.  They 
considered only the $S$-wave interaction and ignored the tensor 
component.  Their potential for the exchange of the $i^{\rm th}$ meson 
was of the form
\begin{equation}
V_i(r) = V^{(i)}_c(r) 
       + \vec{\sigma}_1\cdot\vec{\sigma}_2\ 
       V^{(i)}_\sigma(r)                 \ ,     \label{eq:2.2}
\end{equation}
where the radial potential $V^{(i)}_\alpha, \alpha = c, \sigma$ for 
a meson of mass $m_{i}$ was assumed to be
\begin{equation}
V_{\alpha}^{(i)}(r) = V_0^{(i)} \left[\frac{e^{-m_ir}}{m_ir} 
          - C\,\left(\frac{M}{m_i}\right)\, 
                     \frac{e^{-M r}}{M r}\right]
            \ ;\ \ \alpha = c,\sigma\ .            \   \label{eq:2.3}
\end{equation}
\begin{figure}
\centering\includegraphics[scale=0.48]{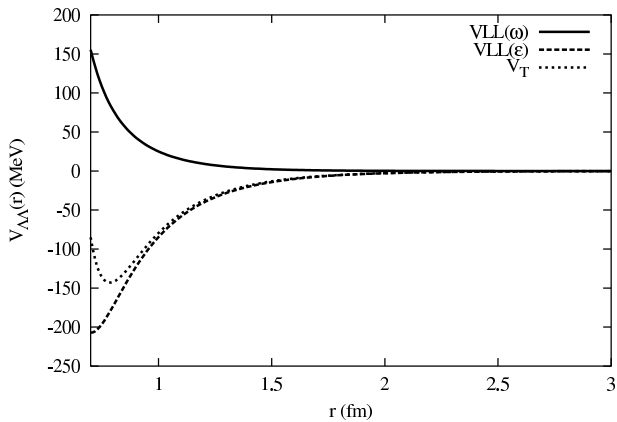}
\caption{The $\Lambda\Lambda$ potential in the $^1$S$_0$ channel. The 
solid and dashed lines indicate the contributions of the $\omega$ and 
$\varepsilon$ exchange, while the dotted line is the total potential. 
$C$ was adjusted so that the $\Lambda\Lambda$ scattering length is
$a_{\Lambda\Lambda}=-1.91$~fm.}\label{fig1}
\end{figure}

To guarantee a one parameter short range repulsion, the mass 
$M=2500$~MeV was used in all partial waves.  Then the remaining
parameter $C$ determined the strength of the short range interaction. 
This new parameter $C$ was constrained to ensure that the potential 
for $r\ge 1.0$~fm is unchanged and that the short range interaction 
is always repulsive.  In Fig.~\ref{fig1} we illustrate the 
$\Lambda\Lambda$ potential in the $^1$S$_0$ channel.  Included in 
the figure are the contributions from the $\varepsilon$ (dashed line) 
and $\omega$ (solid line) exchange as well as the full potential, 
which includes the sum of contributions from all the allowed 
meson exchanges.  In this case the parameter $C$ was adjusted so
that the potential gives a $\Lambda\Lambda$ scattering length 
$a_{\Lambda\Lambda}=-1.91$~fm.  We note that the dominant contribution 
to the potential is from $\varepsilon$ exchange, which is not part of any 
meson octet and was introduced to give medium range attraction and 
to emulate two-pion exchange.

\begin{figure}
\centering\includegraphics[scale=0.48]{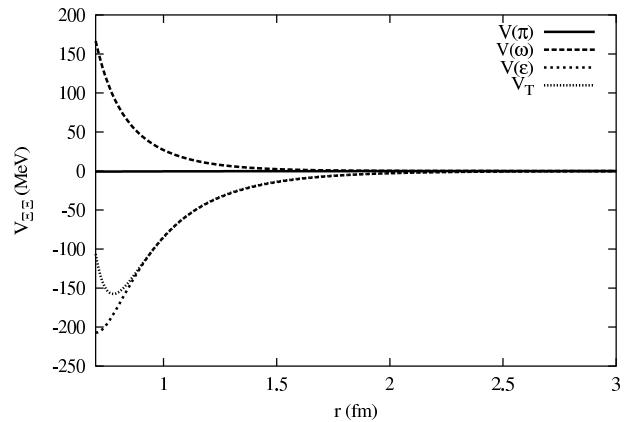}
\caption{The $^1$S$_0$ N$\Xi$-N$\Xi$ potential. The contributions of
the $\pi$, $\omega$ and $\varepsilon$ exchange are represented by
solid, dashed and dotted lines respectively. The total potential is
represented by a dense dotted line.  $C$ was adjusted to give 
$a_{\Lambda\Lambda}=-1.91$~fm.}\label{fig2}
\end{figure}

We now turn to the N$\Xi$-N$\Xi$ potential where $\pi$ exchange is 
allowed.  
In Fig.~\ref{fig2} we present the most important contributions to 
the potential as well as the contribution from $\pi$ exchange.  
Surprisingly, $\pi$ exchange is negligible, as again 
the dominant contribution is from $\varepsilon$ exchange.  One can 
make the same observation for the $\Sigma\Sigma$-$\Sigma\Sigma$ 
potential where $\pi$ exchange is an order of magnitude smaller than 
$\varepsilon$ exchange.  This is a reflection of the fact that in the 
$^1$S$_0$ channel the strength of the $\pi$ exchange includes a factor 
of $\frac{m_\pi}{m}$, where $m$ is a hadron mass.  Thus, we conclude 
that the diagonal elements of the potential contain little 
contribution from $\pi$ exchange and are dominated by $\varepsilon$ exchange. 
If one examines the coupling between the three channels $\Lambda\Lambda$, 
N$\Xi$, and $\Sigma\Sigma$, one observes that $\pi$ exchange 
contributes to the transition between the $\Lambda\Lambda$ and 
$\Sigma\Sigma$ channels.  However, in this case the other isovector 
exchange, the $\rho$, is dominant.

Continuing the analysis, we consider the importance of the coupling 
between the three channels in our OBE Model $D$ based approach. 
We should point out that if the coupling is important, then the 
extraction of the $\Lambda\Lambda$ interaction from light 
$S=-2$ hypernuclei will require that we include this coupling in the 
analysis of the data. To illustrate this point let us consider the 
effective matrix element for the $\Lambda\Lambda$ interaction in 
second order in perturbation theory, \textit{i.e.}
\begin{equation}
V^{\rm eff}_{\Lambda\Lambda}\approx \bra\Lambda\Lambda|V
|\Lambda\Lambda\ket-\frac{|\bra\Lambda\Lambda|V
|N \Xi\ket|^2}{\Delta E}    \ ,                            \label{eq:2.4}
\end{equation}
where $\Delta E\approx 25$~MeV.  In free space due to the small
difference between the $\Lambda\Lambda$ and N$\Xi$ threshold this 
coupling is more important than that between the NN and N$\Delta$ 
in the $S=0$ channel.  On the other hand, in the nuclear medium, the 
transition from $\Lambda\Lambda$ to N$\Xi$ is Pauli blocked.  As a 
result the additional attraction from the second order term is 
suppressed in nuclei.  This implies that the effective 
$\Lambda\Lambda$ matrix element should be less attractive in the 
nuclear medium than in free space.  This is true provided the 
coupling is, in general, large in free space.  Therefore, we 
consider the effective role of the coupling as the size of the
$\Lambda\Lambda$  scattering length $a_{\Lambda\Lambda}$ is 
changed.

\begin{figure}
\centering\includegraphics[scale=0.48]{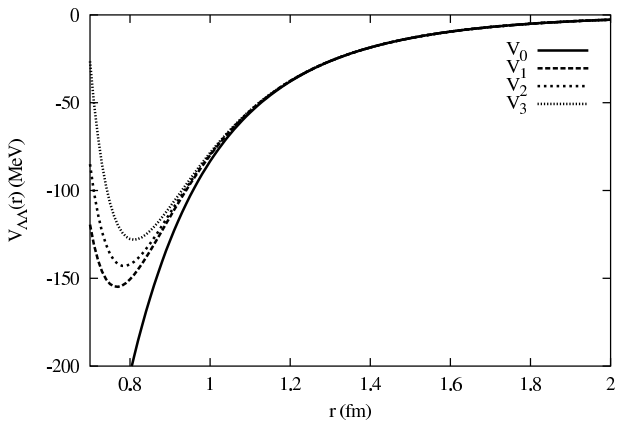}
\caption{The $\Lambda\Lambda$ potential in the $^1$S$_0$ channel. 
The solid line labelled $V_0$ is the OBE with no cutoff. The curves 
$V_1$, $V_2$ and $V_3$ correspond to the potential with no channel 
coupling, with coupling to the N$\Xi$ channel only, and with the full 
coupling to the N$\Xi$ and $\Sigma\Sigma$ channels. The parameter 
$C$ was adjusted to obtain a scattering length 
$a_{\Lambda\Lambda}=-1.91$~fm.}\label{fig3}
\end{figure}

In Fig.~\ref{fig3} we present the $\Lambda\Lambda$ potential with 
no coupling ($V_1$), with coupling to the N$\Xi$ channel ($V_2$), and 
with the full coupling to both N$\Xi$ and $\Sigma\Sigma$ channels 
($V_3$). The short range parameter $C$ was adjusted so that in
each case the potential has a scattering length 
$a_{\Lambda\Lambda}=-1.91$~fm. This potential gives a 
$^{\ \ 6}_{\Lambda\Lambda}$He\cite{C97} binding energy of some 10 MeV 
in the case of $V_1$ and about 9.7 MeV in the case of $V_2$, which 
are somewhat smaller than the experimental result (10.9 MeV) of 
Prowse\cite{P66}.  From the figure we observe that as one includes 
first the N$\Xi$ and then the $\Sigma\Sigma$ channel, the 
$\Lambda\Lambda$ potential becomes shallower.  This suggests that 
the coupling will reduce the binding energy of $\Lambda\Lambda$ 
hypernuclei, in agreement with the result observed by 
Carr~\textit{et al.}\cite{C97}.   Surprisingly, the coupling to 
the $\Sigma\Sigma$ channel is quite important, even though the 
threshold for the $\Sigma\Sigma$ channel is some 160~MeV above the 
$\Lambda\Lambda$ threshold.  One would, therefore, anticipate that 
a free space $\Lambda\Lambda$ interaction somewhat stronger than 
the one considered with $a_{\Lambda\Lambda}=-1.91$~fm would be 
required to reproduce the Prowse datum. 

\begin{figure}
\centering\includegraphics[scale=0.48]{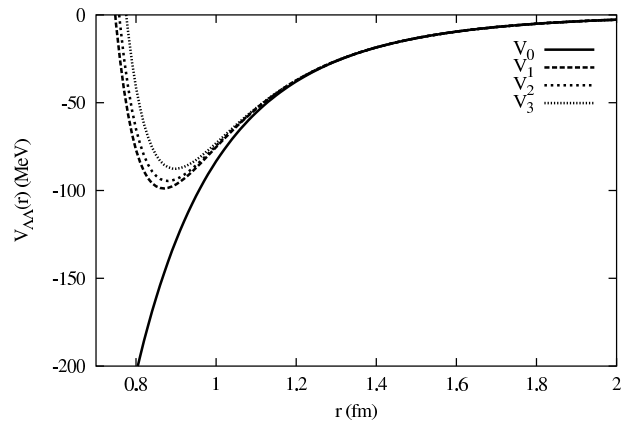}
\caption{The $^1$S$_0$ $\Lambda\Lambda$ potential for the case when 
the potential, including coupling to all the channels gives a
scattering length of $a_{\Lambda\Lambda}=-0.5$~fm. The curves have 
the same labeling as in Figure~\ref{fig3}.}\label{fig4}
\end{figure}

In contrast, the new measurement of the $\Lambda\Lambda$ binding
energy in $^{\ \ 6}_{\Lambda\Lambda}$He\cite{T01} suggests that 
the $\Lambda\Lambda$ potential is, in fact, much weaker than implied
by the earlier measurement.  We, therefore, have considered a 
potential that gives a scattering length $a_{\Lambda\Lambda}=-0.5$~fm. 
This is consistent with the results for the later Nijmegen 
soft core potential\cite{S99}.  In Fig.~\ref{fig4} we present the 
$\Lambda\Lambda$ potential with no coupling ($V_1$), with coupling 
to the N$\Xi$ channel ($V_2$), and with coupling to both the N$\Xi$ 
and $\Sigma\Sigma$ channels ($V_3$) for $a_{\Lambda\Lambda}=-0.5$~fm.
There are two distinct differences between the results for 
$a_{\Lambda\Lambda}=-0.5$~fm and those for 
$a_{\Lambda\Lambda}=-1.91$~fm.  These are: 
(i)~In general the smaller scattering length gives a potential that 
is 30\% shallower. 
(ii)~Of more significance is the fact that the importance of the 
coupling is reduced.  (However, even in this case, the coupling to 
the $\Sigma\Sigma$ channel is more important than just including the 
coupling to the N$\Xi$ channel.)  This suggests that as we reduce the 
strength of the $\Lambda\Lambda$ interaction in our OBE Model $D$ 
based potential, the role of the coupling is reduced.  Perhaps 
more important is the distinct possibility that we may need to 
include the coupling to the $\Sigma\Sigma$ channel even though the 
$\Sigma\Sigma$ threshold is some 160~MeV above that of 
$\Lambda\Lambda$ channel.

\begin{table}
\caption{Variation in the $\Lambda\Lambda$ interaction with changes 
in the strength of the $\Lambda\Lambda$ potential as measured by 
$a_{\Lambda\Lambda}$.}
\vspace{0.3cm}
{\footnotesize
\begin{tabular}{|c|c|c|c|}
\hline 
$a_{\Lambda\Lambda}$ & BE($\Lambda\Lambda\alpha$-N$\Xi\alpha$) &
BE($\Lambda\Lambda\alpha$) & $\Delta B$ \\ 
(fm) & (MeV) & (Mev) & (MeV) \\
\hline
-1.91 & 9.738 & 10.007 & 3.60 \\
-21.1 & 12.268 & 14.138 & 6.13 \\
7.82 & 15.912 & 17.842 & 9.77 \\
3.37 & 19.836 & 23.342 & 13.70 \\ 
\hline
\end{tabular}\label{table1}}
\end{table}

To give some quantitative measure to the variation in the 
$\Lambda\Lambda$ matrix element with changes in the scattering 
length, we recall the results of Ref.\cite{C97} in 
Table~\ref{table1} for $^{\ \ 6}_{\Lambda\Lambda}$He.  Here we
tabulate the $\Lambda\Lambda$ scattering length $a_{\Lambda\Lambda}$, 
and the binding energy of $^{\ \ 6}_{\Lambda\Lambda}$He with and 
without coupling between the $\Lambda\Lambda$ and N$\Xi$ channels.  
Also included are the $\Lambda\Lambda$ matrix elements 
\{$\Delta B=$ [BE$(\Lambda\Lambda$-N$\Xi$)  - 6.14] 
$\approx - |\bra\Lambda \Lambda|V|\Lambda\Lambda\ket| $\} in this 
hypernucleus.  These results confirm our expectation that the 
coupling between the channels becomes weaker in our OBE potential
(based upon the Nijmegen Model $D$) as the scattering length 
$a_{\Lambda\Lambda}$ becomes smaller and negative, \textit{i.e.} as 
the $\Lambda\Lambda$ interaction becomes weaker.

\begin{figure}
\centering\includegraphics[scale=0.54]{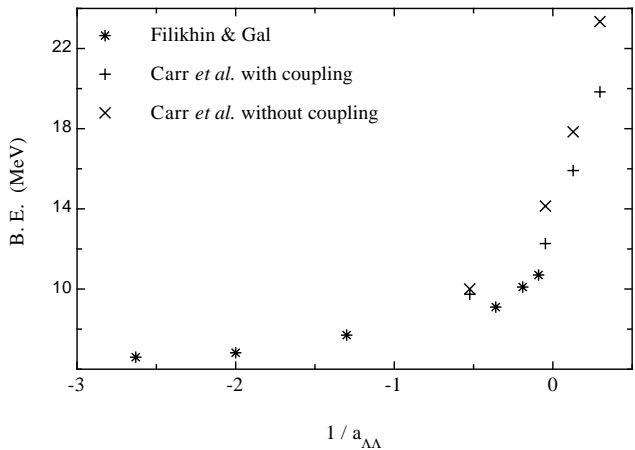}
\caption{Plot of the binding energy (B.E.) of 
$^{\ \ 6}_{\Lambda\Lambda}$He as a function of 
$a^{-1}_{\Lambda\Lambda}$. Here (\textbf{+}) and (\textbf{x}) are 
the results of Carr~\textit{et al.}  with and without coupling to 
the N$\Xi$ channel. Also included are the results of Filikhin and 
Gal ({\bf *}).}\label{fig5}
\end{figure} 

This change in the binding energy, with and without the coupling 
to the N$\Xi$ channel, as one varies the $\Lambda\Lambda$ scattering 
length, is illustrated in Fig.~\ref{fig5}.  Here we plot the binding 
energy as a function of $a^{-1}_{\Lambda\Lambda}$.  In particular, 
the (\textbf{+}) and (\textbf{x}) are the results of Ref.\cite{C97} 
with and without coupling between the $\Lambda\Lambda$ and N$\Xi$ 
channels. Also included are the recent results of Filikhin and 
Gal (FG)\cite{F02} which are calculated with only the $\Lambda\Lambda$ 
channel (\textit{i.e.}, no channel coupling is included).  From the 
results of Ref.\cite{C97} we can clearly see that the role of 
coupling for a small, negative scattering length would be 
negligible, while the results of FG\cite{F02} suggest 
that the new experimental result\cite{T01} for the binding energy 
of $^{\ \ 6}_{\Lambda\Lambda}$He of $7.25\pm0.2^{+0.18}_{-0.11}$~MeV 
will imply a $\Lambda\Lambda$ scattering length of $\approx -0.5$~fm.

In conclusion, we have demonstrated that within the framework of an 
OBE model and flavour $SU(3)$ corresponding to the Nijmegen
Model $D$ one can generate a one parameter set of potentials that 
preserve the OBE tail.  The short range repulsion can then be 
adjusted to give the $\Lambda\Lambda$ scattering length.  The 
primary concern with this procedure is the fact that the potential 
is dominated by the exchange of the scalar $\varepsilon$ meson. This 
meson was introduced in the strangeness $S=0$ sector to give medium 
range attraction and to model two pion exchange.  Its dominance in 
the $S=-2$ channel suggests that one should go back and include 
explicit two-pion exchange within a framework that will still allow 
one to perform a flavour $SU(3)$ rotation of the potential to 
generate the $\Lambda\Lambda$ interaction.  From the analysis of 
the importance of coupling between the ($\Lambda\Lambda$, N$\Xi$, 
and $\Sigma\Sigma$) channels in the strangeness $S=-2$ $^1$S$_0$
partial wave, we found that for a small, negative $\Lambda\Lambda$ 
scattering length the coupling between the channels is relatively 
weak.  If we now combine this observation with the recent 
measurement of the binding energy of $^{\ \ 6}_{\Lambda\Lambda}$He, 
one may conclude that a confirmation of this measurement could 
constrain the $\Lambda\Lambda$ scattering length to 
$a_{\Lambda\Lambda} \approx -0.5$~fm with good accuracy.  Such a 
feable interaction would not require inclusion of the coupling 
to the N$\Xi$ and $\Sigma\Sigma$ channels, which is a complication 
in the calculation of energies of light hypernuclei, if the OBE 
model used here is a valid representation of the physics.  Finally, 
if the new measurement of the $\Lambda\Lambda$ matrix 
element\cite{T01} is correct, then it would confirm the validity 
of the $SU(3)$ prediction for the relative strengths of the 
interactions in the $S=0, -1$, and $-2$ sectors as stated in 
Eq.~(\ref{eq:1.3}).

\section{Acknowledgements}

I.R.A.\ would like to thank the Australian Research Council for 
their financial support during the course of this work.  The 
research of B.F.G. is supported by the U.S.\ Department of Energy 
under contract W-7405-ENG-36.  B.F.G. gratefully acknowledges 
the support of the Alexander von Humboldt-Stiftung. The authors would 
like to thank Professor J. Speth for the 
hospitality of the Institut f\"ur Kernphysik at the 
Forschungszentrum-Juelich.

\bibliography{llpot}

\end{document}